\documentclass[pre,amsfonts,amssymb,amsmath,floatfix,twocolumn]{revtex4}
 \usepackage{graphicx}

\newcommand{\etal}{\textit{ et al. }}
\newcommand{\p}{\partial}
\renewcommand{\vec}[1]{\mathbf{#1}}
\newcommand{\abs}[1]{\left|#1\right|}

\newcommand{\rin}{r_{\textrm{in}}}
\newcommand{\rex}{r_{\textrm{exit}}}
\newcommand{\rout}{r_{\textrm{out}}}

\begin{document}
\title{Analysis of Granular Flow in a Pebble-Bed Nuclear Reactor}
\author{Chris H. Rycroft$^1$, Gary S. Grest$^2$, James W. Landry$^3$, and
Martin Z. Bazant$^1$}
\affiliation{$^1$Department of Mathematics, Massachusetts Institute of
Technology, Cambridge, MA 02139}
\affiliation{$^2$Sandia National Laboratories, Albuquerque, NM 87185}
\affiliation{$^3$Lincoln Laboratory, Massachusetts Institute of Technology,
Lexington, MA 02420}

\begin{abstract}
  Pebble-bed nuclear reactor technology, which is currently being revived
  around the world, raises fundamental questions about dense granular flow in
  silos. A typical reactor core is composed of graphite fuel pebbles, which
  drain very slowly in a continuous refueling process. Pebble flow is poorly
  understood and not easily accessible to experiments, and yet it has a major
  impact on reactor physics. To address this problem, we perform full-scale,
  discrete-element simulations in realistic geometries, with up to 440,000
  frictional, viscoelastic $6\textrm{cm}$-diameter spheres draining in a
  cylindrical vessel of diameter $3.5\textrm{m}$ and height $10\textrm{m}$ with
  bottom funnels angled at 30$^\circ$ or 60$^\circ$. We also simulate a
  bidisperse core with a dynamic central column of smaller graphite moderator
  pebbles and show that little mixing occurs down to a 1:2 diameter ratio. We
  analyze the mean velocity, diffusion and mixing, local ordering and porosity
  (from Voronoi volumes), the residence-time distribution, and the effects of
  wall friction and discuss implications for reactor design and the basic
  physics of granular flow.
\end{abstract}
\maketitle

\section{Introduction}\label{sec:intro}

\subsection{Background}

A worldwide effort is underway to develop more economical, efficient,
proliferation resistant, and safer nuclear power~\cite{nuclearweb}. A promising
Generation IV reactor design is the uranium-based, graphite moderated,
helium-cooled very high temperature reactor~\cite{iaea01}, which offers
meltdown-proof passive safety, convenient long-term waste storage, modular
construction, and a means of nuclear-assisted hydrogen production and
desalination. In one embodiment, uranium dioxide is contained in microspheres
dispersed in spherical graphite pebbles, the size of billiard balls, which are
very slowly cycled through the core in a dense granular
flow~\cite{nicholls01,talbot02}. Control rods are inserted in graphite bricks
of the core vessel, so there are no obstacles to pebble flow.

The pebble-bed reactor (PBR) concept, which originated in Germany in the 1950s,
is being revisited by several countries, notably China~\cite{wired}
(HTR-10~\cite{hu04}) and South Africa~\cite{nicholls01} (PBMR~\cite{pbmr}),
which plan large-scale deployment. In the United States, the Modular Pebble Bed
Reactor (MPBR) ~\cite{talbot02,MPBR} is a candidate for the Next Generation
Nuclear Plant of the Department of Energy. A notable feature of MPBR (also
present in the original South African design) is the introduction of graphite
moderator pebbles, identical to the fuel pebbles but without the uranium
microspheres. The moderator pebbles form a dynamic central column, which serves
to flatten the neutron flux across the annular fuel region without placing any
fixed structures inside the core vessel. The annular fuel region increases the
power output and efficiency, while preserving passive safety. In the bidisperse
MPBR, the moderator pebbles are smaller to reduce the permeability of the
central column and thus focus helium gas on the outer fuel annulus. The
continuous refueling process is a major advantage of pebble-bed reactors over
other core designs, which typically require shutting down for a costly
dismantling and reconstruction. The random cycling of pebbles through a flowing
core also greatly improves the uniformity of fuel burnup.

In spite of these advantages, however, the dynamic core of a PBR is also a
cause for concern among designers and regulators, since the basic physics of
dense granular flow is not fully understood. Indeed, no reliable continuum
model is available to predict the mean velocity in silos of different
shapes~\cite{choiexpt}, although the empirical Kinematic
Model~\cite{mullins72,nedderman79,nedderman} provides a reasonable fit near the
orifice in a wide silo~\cite{tuzun79,medina98a,samadani99,choi04}. A
microscopic model for random-packing dynamics has also been
proposed~\cite{spot-ses} and fitted to reproduce drainage in a wide
silo~\cite{ssim}, but a complete statistical theory of dense granular flow is
still lacking. The classical kinetic theory of gases has been successfully
applied to dilute granular flows~\cite{savage79,jenkins83,prakash91}, in spite
of problems with inelastic collisions~\cite{kadanoff99}, but it clearly breaks
down in dense flows with long-lasting, frictional
contacts~\cite{natarajan95,choi04}, as in pebble-bed reactors. Plasticity
theories from soil mechanics might seem more appropriate~\cite{nedderman}, but
they cannot describe flows in silos of arbitrary shape and often lead to
violent instabilities~\cite{schaeffer87,pitman87}, although a stochastic flow
rule~\cite{kamrin06} may resolve these difficulties and eventually lead to a
general theory.

For now, experiments provide important, although limited, information about
dense granular flows. Many experiments have been done on drainage flows in
quasi-2d silos where particles are tracked accurately at a transparent
wall~\cite{medina98a,medina98b,samadani99,choi04,choiexpt}. Some
three-dimensional particle tracking in granular materials and colloids has also
been done with magnetic resonance imaging~\cite{mueth00}, confocal
microscopy~\cite{weeks00}, index matching with an interstitial
fluid~\cite{tsai04}, and diffusing-wave spectroscopy~\cite{menon97}, although
these systems are quite different from a pebble-bed reactor core. Experimental
studies of more realistic geometries for PBR have mostly focused on the
porosity distribution of static packings of
spheres~\cite{goodling83,sederman01}, which affects helium gas flow through the
core~\cite{cohen81,vortmeyer83,white87}.

As a first attempt to observe pebble dynamics experimentally in a reactor
model, the slow flow of plastic beads has recently been studied in 1:10 scale
models of MPBR in two different ways~\cite{kadak04}: The trajectories of
colored pebbles were recorded (by hand) along a plexiglass wall in a half-core
model, and a single radioactive tracer pebble in the bulk was tracked in three
dimensions in a full-core model. Very slow flow was achieved using a screw
mechanism at the orifice to approximate the mean exit rate of one pebble per
minute in MPBR. These experiments demonstrate the feasibility of the dynamic
central column and confirm that pebbles diffuse less than one diameter away
from streamlines of the mean flow. However, it is important to gain a more
detailed understanding of pebble flow in the entire core to reliably predict
reactor power output, fuel efficiency, power peaking, accident scenarios using
existing nuclear engineering codes~\cite{terry02,gougar02}.

\subsection{Discrete-Element Simulations}

Simulations are ideally suited to provide complete, three-dimensional
information in a granular flow. Some simulations of the static random packing
of fuel pebbles in a PBR core have been reported~\cite{dutoit02,ougouag05}, but
in the last few years, large-scale, parallel computing technology has advanced
to the stage where it is now possible to carry out simulations of continuous
pebble flow in a full-sized reactor geometry using the Discrete Element Method
(DEM). In such simulations, each particle is accurately modeled as a sphere
undergoing realistic frictional interactions with other particles
\cite{silbert01,landry03}. In this paper, we present DEM simulations which
address various outstanding issues in reactor design, such as the sharpness of
the interface between fuel and moderator pebbles (in both monodisperse and
bidisperse cores), the horizontal diffusion of the pebbles, the geometry
dependence of the mean streamlines, the porosity distribution, wall effects,
and residence-time distributions.

\begin{figure}
	\centering
	\includegraphics[width=3.8cm]{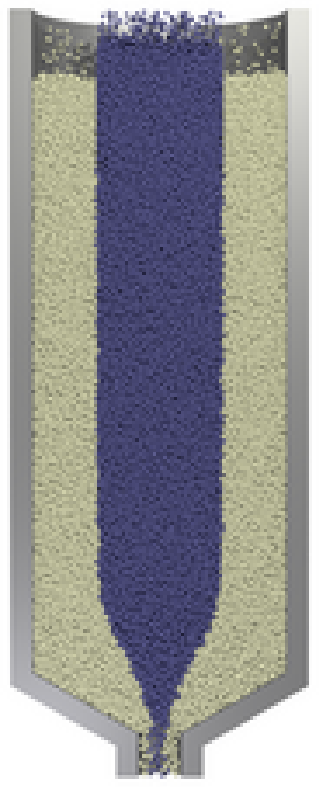}
	\qquad
	\includegraphics[width=3.8cm]{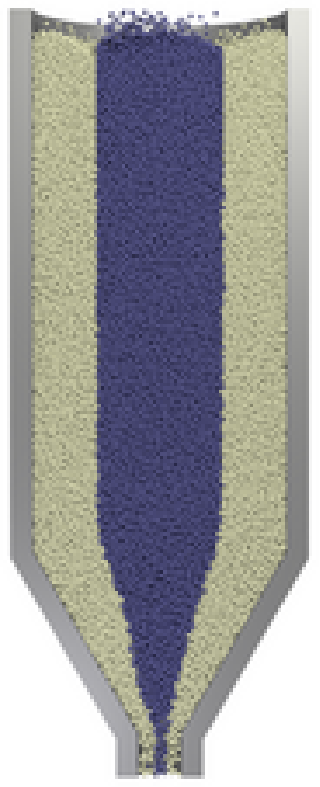}
	\caption{(Color online) Snapshots of vertical cross-sections of the
	simulations for the two geometries considered in this report. We make
	use of a cylindrical coordinate system $(r,\theta,z)$ where $z=0$ at
	the orifice. At the base of the container there is a small exit pipe of
	radius $\rex=5d$ that extends upwards to $z=10d$. This connects to a
	conical funnel region, which has slope thirty degrees (left) or sixty
	degrees (right). The conical wall connects to a cylindrical wall of
	radius $\rout=29d$, at $z=23.86d$ and $z=51.57d$ for the thirty and
	sixty degree reactor geometries respectively. Particles are poured into
	the container up to a height of approximately $z=160d$. A cylindrical
	wall at $\rin=14.5d$ extends down into the packing to a height of
	$z=140d$ to keep the two types of pebbles mixing at the surface.}
	\label{diag}
\end{figure}

Our simulations are based on the MPBR geometry~\cite{talbot02,MPBR}, consisting
of spherical pebbles with diameter $d=6\textrm{cm}$ in a cylindrical container
approximately $10\textrm{m}$ high and $3.5\textrm{m}$ across. In this design
there is a central column of moderating reflector pebbles, surrounded by an
annulus of fuel pebbles. The two pebble types are physically identical except
that the fuel pebbles contain sand-sized uranium fuel particles. Particles are
continuously cycled, so that those exiting the container are reintroduced at
the top of the packing. In order to efficiently maintain the central column, a
cylindrical guide ring of radius $\rin = 14.5d$ extends into the packing to
$z=140d$. Reflector pebbles are poured inside, while fuel pebbles are poured
outside, and the guide ring ensures that two types do not mix together at the
surface. Figure \ref{diag} shows the two main geometries that were considered;
for much of this analysis, we have concentrated on the case when the exit
funnel is sloped at thirty degrees, but since this angle can have a large
effect on the pebble flow, we also consider the case of the when the funnel is
sloped at sixty degrees. In both cases the radius of the opening at the bottom
of the funnel is $\rex=5d$.

In MPBR, as in most pebble-bed reactors, the drainage process takes place
extremely slowly. Pebbles are individually removed from the base of the reactor
using a screw mechanism, at a typical rate of one pebble per minute, and the
mean residence time of a pebble is 77 days. Carrying out a DEM simulation at
this flow rate would make it infeasible to collect enough meaningful data.
However, previous experimental work by Choi\etal\cite{choi04} has shown that
the regime of slow, dense granular flow is governed by a distinctly non-thermal
picture, where particles undergo long-lasting contacts with their neighbors,
and the features of the flow are predominately governed by geometry and packing
constraints. In particular, they observed that for a large range of hopper
drainage experiments, altering the orifice size resulted in a change in the
overall flow rate, but did not alter the geometry of the flow profile -- the
flow velocities were scaled by a constant factor. Furthermore, geometric
properties of the flow, such as particle diffusion, were unaffected by the
overall flow rate. We therefore chose to study a faster flow regime in which
pebbles drain from the reactor exit pipe under gravity. Our results can be
related directly to the reactor design by rescaling the time by an appropriate
factor.

As well as the two full-scale simulations described above, we also considered a
half-size geometry in order to investigate how various alterations in the
makeup of the reactor would affect the flow. In particular, we examined a
series of bidisperse simulations, in which the diameter of moderator particles
in the central column was reduced. As explained in section~\ref{sec:bi}, this
has the effect of reducing the gas permeability of the central column, thus
focusing the helium coolant flow on the hottest region of the reactor core, in
and around the fuel annulus. The purpose of the simulations is to test the
feasibility of the bidisperse PBR concept, as a function of the size ratio of
moderator and fuel pebbles, with regard to the granular flow. It is not clear
{\it a priori} under what conditions the dynamic column will remain stable with
little interdiffusion of moderator and graphite pebbles. 

To study this issue, we made a sequence of three runs using a half-size reactor
geometry. (The smaller core size is needed since the number of smaller pebbles
increases as the inverse cube of the diameter ratio.) The geometry is similar to
that used above, except that the radius of the cylindrical container is
decreased to $15d$, with the guide ring at $\rin=7.5d$. The radius of the exit
pipe is decreased to $\rex=4d$. In the experiments, we keep the diameter of the
fuel pebbles fixed at $d$, and use $d$, $0.8d$, and $0.5d$ for the diameters of
the moderator pebbles. The same geometry was also used to study the effect of
wall friction, by making an additional run with the particle/wall friction
coefficient $\mu_w=0$.

The paper is organized as follows. In section \ref{sec:model}, we discuss the
simulation technique that was used and briefly describe its implementation.
This is followed with some basic analysis of the velocity profiles and a
comparison to the Kinematic Model in section \ref{sec:vel}. We study diffusion
around streamlines in section \ref{sec:diff} and the distribution of porosity
and local ordering in section \ref{sec:ordering}. Next, in section
\ref{sec:wait} we examine the residence-time distribution of pebbles in the
reactor, which is related to fuel burnup, and in section \ref{sec:fric} we show
that wall friction plays an important role. In section \ref{sec:bi} we analyze
the bidisperse PBR concept with half-size reactor simulations for a range of
pebble-diameter ratios, focusing on the mean flow, diffusion, and mixing. We
conclude in section \ref{sec:conc} by summarizing implications of our study for
reactor design and the basic physics of granular flow.

\section{Models and Methods}\label{sec:model}
The DEM simulations are based on a modified version of the model developed by
Cundall and Strack \cite{cundall79} to model cohesionless particulates
\cite{silbert01,landry03}. Monodisperse spheres with diameter $d$ interact
according to Hertzian, history dependent contact forces. For a distance
$\vec{r}$ between a particle and its neighbor, when the particles are in
compression, so that $\delta=d-\abs{\vec{r}}>0$, then the two particles
experience a force $\vec{F}=\vec{F}_n+\vec{F}_t$, where the normal and
tangential components are given by
\begin{eqnarray}
\vec{F}_n&=&\sqrt{\delta/d} \left(k_n\delta\vec{n} - \frac{\gamma_n\vec{v}_n}{2}\right)\\
\vec{F}_t&=&\sqrt{\delta/d} \left(-k_t\Delta\vec{s}_t -\frac{\gamma_t\vec{v}_t}{2}\right).
\end{eqnarray}
Here, $\vec{n}=\vec{r}/\abs{\vec{r}}$. $\vec{v}_n$ and $\vec{v}_t$ are the
normal and tangential components of the relative surface velocity, and
$k_{n,t}$ and $\gamma_{n,t}$ are the elastic and viscoelastic constants,
respectively. $\Delta\vec{s}_t$ is the elastic tangential displacement between
spheres, obtained by integrating tangential relative velocities during elastic
deformation for the lifetime of the contact, and is truncated as necessary to
satisfy a local Coulomb yield criterion $\abs{\vec{F}_t}\le \mu
\abs{\vec{F}_n}$. Particle-wall interactions are treated identically, though
the particle-wall friction coefficient $\mu_w$ is set independently.

For the monodispersed system, the spheres have diameter $d=6\textrm{cm}$, mass
$m=210\textrm{g}$ and interparticle friction coefficient $\mu=0.7$, flowing under the
influence of gravity $g=9.81\textrm{ms}^{-1}$. For the bi-dispersed systems,
the moderator particles have diameter $0.8d$ or $0.5d$. The particle-wall
friction coefficient $\mu_w=0.7$ except in one case where we model a
frictionless wall, $\mu_w=0.0$. For the current simulations we set
$k_t=\frac{2}{7}k_n$, and choose $k_n=2\times10^5 mg/d$. While this is
significantly less than would be realistic for graphite pebbles, where we
expect $k_n > 10^{10} mg/d$, such a spring constant would be prohibitively
computationally expensive, as the time step scales as $\delta t \propto
k_n^{-1/2}$ for collisions to be modeled effectively. Previous simulations
have shown that increasing $k_n$ does not significantly alter physical results
\cite{landry03}. We use a time step of $\delta t=1.0\times10^{-4}\tau$ and
damping coefficients $\gamma_n=50\tau^{-1}$ and $\gamma_t=0.0$, where
$\tau=\sqrt{d/g}=0.078\textrm{s}$. All measurements are expressed in terms of
$d$, $m$ and $\tau$.

The initial configurations are made by extending the inner cylinder from $140d$
to the bottom of the container, adding a wall at the bottom of the container to
stop particles from draining, and pouring in moderator pebbles into the inner
cylinder and fuel pebbles between the inner and outer cylinders until the
reactor was loaded. The bottom wall is then removed, the inner cylinder is
raised to $140d$, and particles are allowed to drain out of the container. As
noted above, particles are recycled with moderator particles reinserted within
the inner cylinder, and fuel particles between the inner and outer cylinders.
All results presented here are after all the particles have cycled through the
reactor at least once. The number of moderator and fuel particles was adjusted
slightly from the initial filling so that the level at the top of the reactor
is approximately equal. For the full scale simulation with a thirty degree
outlet, the total number of pebbles is 440,000 with 105,011 moderator pebbles
and 334,989 fuel pebbles, while for the sixty degree outlet, the total number
of pebbles is 406,405 with 97,463 moderator and 308,942 fuel pebbles. For the
former case, a million steps took approximately 13 hours on 60 processors on
Sandia's Intel Xenon cluster.

For the bidispersed simulations the total number of pebbles is 130,044,
160,423, and 337,715 for the diameter of the moderator particles equal to $d$,
$0.8d$ and $0.5d$ respectively. As the diameter of the moderator pebbles is
decreased the number of particles required rapidly increases, since it scales
according to the inverse of the diameter cubed.

A snapshot of all the particle positions is recorded every $5\tau =
0.39\textrm{s}$. For the thirty degree reactor geometry we collected 1,087
successive snapshots, totaling 24.9Gb of data, while for the sixty degree
reactor geometry, we collected 881 successive snapshots, totaling 18.7Gb of
data. A variety of analysis codes written in Perl and C++ were used to
sequentially parse the snapshot files to investigate different aspects of the
flow. We also created extended data sets, with an additional 440 snapshots
for the thirty degree geometry, and 368 snapshots for the sixty degree
geometry, for examining long residence times in section \ref{sec:wait}.

\section{Mean-Velocity Profiles}\label{sec:vel}

\subsection{Simulation Results}

Since we have a massive amount of precise data about the positions of the
pebbles, it is possible to reconstruct the mean flow in the reactor with great
accuracy. However care must be taken when calculating velocity profiles to
ensure the highest accuracy. Initial studies of the data showed that
crystallization effects near the wall can create features in the velocity
profile at a sub-particle level, and we therefore chose a method that could
resolve this.

By exploiting the axial symmetry of the system, one only need to find the
velocity profile as a function of $r$ and $z$ only. The container is divided
into bins and the mean velocity is determined within each. A particle which is
at $\vec{x}_n$ at the $n$th timestep and at $\vec{x}_{n+1}$ at the $(n+1)$th
timestep, makes a velocity contribution of $(\vec{x}_{n+1}-\vec{x}_n)/\Delta t$
in the bin which contains its midpoint, $(\vec{x}_{n+1}+\vec{x}_n)/2$.

In the $z$ direction, we divide the container into strips $1d$ across. However,
in the $r$ direction we take an alternative approach. Since the number of
pebbles between a radius of $r$ and $r+\Delta r$ is proportional to $r \Delta
r$, dividing the container into bins of a fixed width is unsatisfactory, since
the amount of data in bins with high $r$ would be disproportionately large. We
therefore introduce a new coordinate $s=r^2$. The coordinate $s$ covers the
range $0<s<\rout^2$, and we divide the container into regions that are equally
spaced in $s$, of width $1d^2$. The number of pebbles in each bin is therefore
roughly equal, allowing for accurate averaging in the bulk and high resolution
at the boundary.

This result yields extremely accurate velocity profiles in the cylindrical
region of the tank. However, it fails to capture crystallization effects in the
conical region: since the particles are aligned with the slope of the walls are
averaged over a strip in $z$ of width $1d$, any effects are smeared out across
several bins. We therefore scaled the radial coordinate to what it would be if
the particle was in the center of the strip. Specifically, if the radius of the
container is given by $R(z)$, a particle at $(r_n,z_n)$ is recorded as having
radial coordinate $r_n R(z)/R(z_n)$. In the cylindrical region of the tank this
has no effect, while in the conical region, it effectively creates
trapezoid-shaped bins from which it is easy to see crystallization effects
which are aligned with the wall.

\begin{figure}
	\input{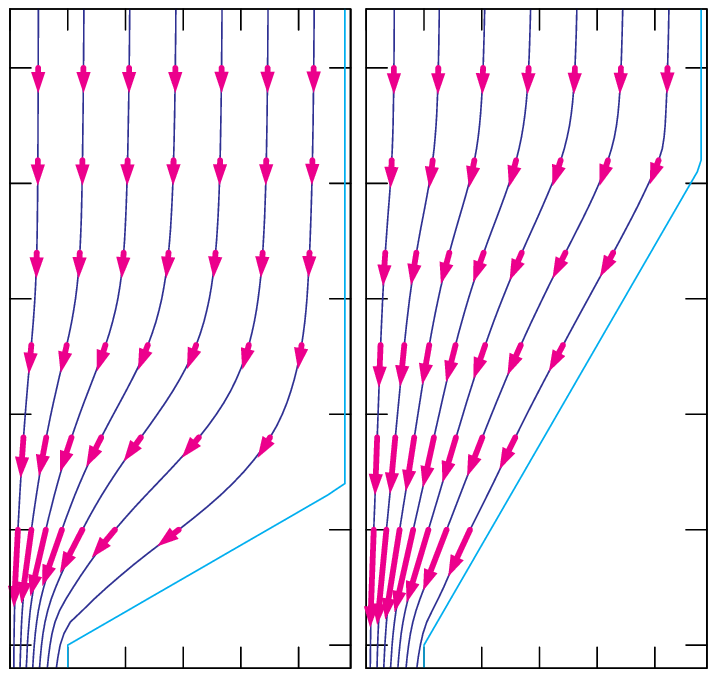}
	\caption{(Color online) Computed streamlines of the mean flow in the
	30$^\circ$ (left) and 60$^\circ$ reactor geometries. Arrows are
	proportional to the velocity vectors in selected horizontal slices.}
	\label{stream}
\end{figure}

The streamlines of the mean flow are shown in Fig.~\ref{stream} in the two
geometries. Streamlines are computed by Lagrangian integration of the DEM
velocity field, starting from points at a given height, equally spaced in
radius. In each geometry, there is a transition from a nonuniform converging
flow in the lower funnel region to a nearly uniform plug flow in the upper
cylindrical region, consistent with the standard engineering picture of silo
drainage~\cite{nedderman}. In the wider funnel, there is a region of much
slower flow near the sharp corner at the upper edge of the funnel. Our results
for both geometries are quite consistent with particle-tracking data for
quasi-2d silos of similar shapes~\cite{choiexpt} and half-cylinder models of
the MPBR core~\cite{kadak04}, which provides an important validation of our
simulations.

\begin{figure}
	\centering
	\input{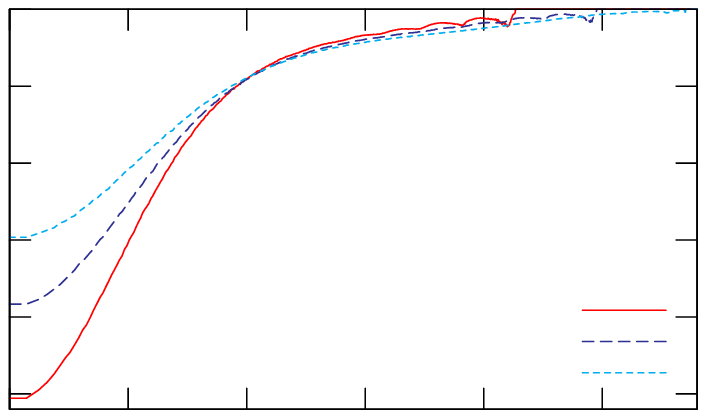}
	\input{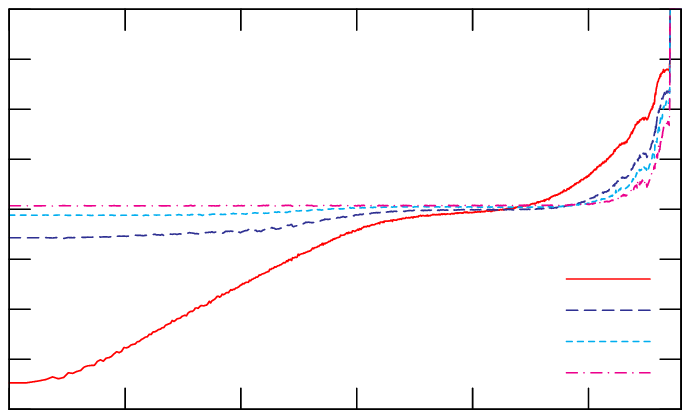}
	\caption{(Color online) Velocity profiles for the thirty degree reactor
	geometry for several low cross-sections (a) and several high
	cross-sections (b).}
	\label{v30}
\end{figure}

We now look more closely at horizontal slices of the velocity field. Figure
\ref{v30}(a) shows several velocity profiles for the thirty degree case in the
narrowing section of the container. As expected, we see a widening of the
velocity profile as $z$ increases. We can also see lattice effects, spaced at
$\sqrt{3}d$ apart, due to to particles crystallizing on the conical wall
section.

Figure \ref{v30}(b) shows similar plots for several heights in the upper region
of the container. At these heights, the velocity profile is roughly uniform
across the container. However a boundary layer of slower velocities, several
particle diameters wide, still persists. The average velocities of particles
touching the boundary is between one half and two thirds that of particles in
the bulk; it is expected that this behavior is very dependent on particle-wall
friction; this issue is studied in more detail in section \ref{sec:fric}.

High in the container, results for the sixty degree geometry are very similar
to the thirty degree case (and thus are not shown). However, as would be
expected, a significantly different crossover from parabolic flow to plug-like
flow in the lower part of the tank is observed, as shown in figure
\ref{kincomp2}.


\subsection{Comparison with the Kinematic Model}\label{subsec:kin}
Perhaps the only continuum theory available for the mean flow profile in a
slowly draining silo is the Kinematic
Model~\cite{lit58,mullins72,nedderman79,nedderman}, which postulates that
horizontal velocity vector $\vec{u}$ is proportional to the horizontal gradient
$\nabla_\perp$ of the downward velocity component $v$ (i.e. the local shear
rate),
\begin{equation}
\vec{u} = b\nabla_\perp v, \label{eq:u}
\end{equation}
where $b$ is the ``diffusion length'', a material parameter typically in the
range of one to three particle diameters. The idea behind Eq.~(\ref{eq:u}) is
that particles drift from regions of low to high downward velocity, where there
are more local rearrangements (and more free volume) to accommodate their
collective motion. The approximation of incompressibility,
$\nabla\cdot(\vec{u},-v)=0$, applied to Eq.~(\ref{eq:u}) yields a diffusion
equation for the downward velocity,
\begin{equation}
	\frac{\p v}{\p z} = b \nabla_\perp^2 v,  \label{eq:veq}
\end{equation}
where the vertical coordinate $z$ acts like ``time''. Boundary conditions on
Eq.~(\ref{eq:veq}) require no normal velocity component at the container walls,
except at the orifice, where $v$ is specified (effectively an ``initial
condition''). As described in Appendix \ref{appendix_kin}, this boundary-value
problem can be accurately solved using a standard Crank-Nicholson scheme for
the diffusion equation.

\begin{figure}
	\input{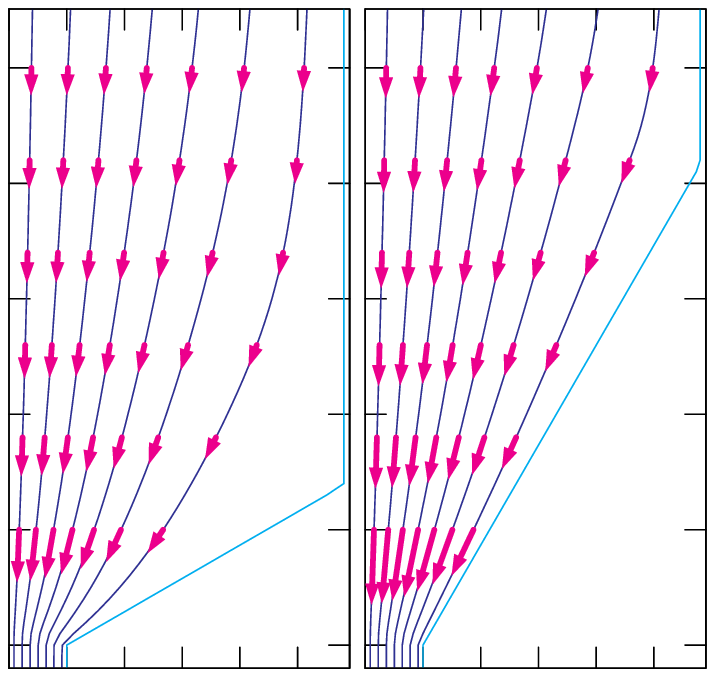}
	\caption{(Color online) Streamlines of the mean flow in the 30$^\circ$
	(left) and 60$^\circ$ reactor geometries for the numerical solution of
	the Kinematic Model. Arrows are proportional to the velocity vectors in
	selected horizontal slices.}
	\label{stream_kin}
\end{figure}

The kinematic parameter $b$ can be understood as a diffusion length for free
volume, which is introduced at the orifice and diffuses upward, causing
downward diffusion of particles. It was originally proposed that free volume is
carried by voids~\cite{lit58,mullins72}, which displace single particles as
they move, but a more realistic mechanism involves cooperative particle motion
due to diffusing ``spots'' of delocalized free volume~\cite{spot-ses}. The Spot
Model can produce accurate flowing packings in wide silos~\cite{ssim}, and the
Kinematic Model can be derived as the continuum limit of the simplest case
where spots drift upward at constant velocity (due to gravity) while undergoing
independent random walks, although more general continuum equations are also
possible for different spot dynamics. A first-principles mechanical theory of
spot dynamics is still lacking (although it may be based on a stochastic
reformulation of Mohr-Coulomb plasticity~\cite{kamrin06}), so here we simply
try a range of $b$ values and compare to the DEM flow profiles.

\begin{figure}
	\centering
	\input{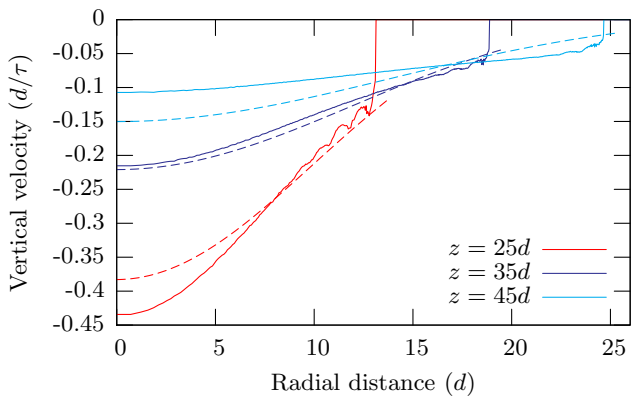}
	\caption{(Color online) Velocity profiles for the $60^\circ$ reactor
	geometry (solid lines), with a comparison to the Kinematic Model for
	$b=3d$ (dashed lines).}
	\label{kincomp2}
\end{figure}

Consistent with a recent experimental study of quasi-2d silos~\cite{choiexpt},
we find reasonable agreement between the Kinematic Model predictions and the
DEM flow profiles, but the effect of the container geometry is not fully
captured. In the converging flow of the funnel region, the streamlines are
roughly parabolic, as predicted by the Kinematic Model and found in many
experiments~\cite{tuzun79,medina98a,samadani99,choi04,choiexpt}. For that
region, it is possible to choose a single value ($b=3d$) to achieve an
acceptable fit to the DEM flow profiles for both the $30^\circ$ and $60^\circ$
funnel geometries, as shown in figure \ref{kincomp2}. 

In spite of the reasonable overall fit, the Kinematic Model has some problems
describing the DEM results. It fails to describe the several particle thick
boundary layer of slower velocities seen in the DEM data. In the original
model, $b$ depends only on the properties of the granular material, but we find
that it seems to depend on the geometry; the best fit to the $30^\circ$ DEM
data is $b\approx 2.5d$, while the best fit for the $60^\circ$ DEM data is
$b\approx 3.0d$. Such discrepancies may partly be due to the boundary layers,
since in the lower section of the container the conical walls may have an
appreciable effect on the majority of the flow. We also find that the Kinematic
Model fails to capture the rapid transition from converging flow to plug flow
seen in the DEM data. This is shown clearly by comparing the streamlines for
the Kinematic Model in figure \ref{stream_kin} with those for DEM. Streamlines
for the Kinematic Model are roughly parabolic, and no single value of $b$ can
capture the rapid change from downward streamlines to converging streamlines
seen in DEM.

The difficulty in precisely determining $b$ is also a common theme in
experiments, although recent data suggests that a nonlinear diffusion length
may improve the fit~\cite{choiexpt}. Perhaps a more fundamental problem with
the Kinematic Model is that it cannot easily describe the rapid crossover from
parabolic converging flow to uniform plug flow seen in both geometries our DEM
simulations; we will return to this issue in section \ref{sec:ordering}.

\section{Diffusion and Mixing}
\label{sec:diff}

Nuclear engineering codes for PBR core neutronics typically assume that pebbles
flow in a smooth laminar manner along streamlines, with very little lateral
diffusion~\cite{terry02,gougar02}. Were such significant diffusion to occur
across streamlines, it could alter the core composition in unexpected ways. In
the MPBR design with a dynamic central column~\cite{MPBR}, diffusion leads to
the unwanted mixing of graphite pebbles from the central reflector column with
fuel pebbles from the outer annulus, so it must be quantified. Simulations and
experiments are crucial, since diffusion in slow, dense granular flows is not
fully understood~\cite{spot-ses}.

Particle-tracking experiments on quasi-2d silos~\cite{choi04} and half-cylinder
MPBR models~\cite{kadak04} have demonstrated very little pebble diffusion in
slow, dense flows, but the observations were made near transparent walls, which
could affect the flow, e.g. due to ordering (see below). Three-dimensional
tracking of a radioactive tracer in a cylindrical MPBR model has also shown
very little diffusion, at the scale of a single pebble diameter for the
duration of the flow~\cite{kadak04}. Here, we take advantage of the complete
information on pebble positions in our DEM simulations to study core diffusion
and mixing with great accuracy.

\begin{figure}
	\centering
	\input{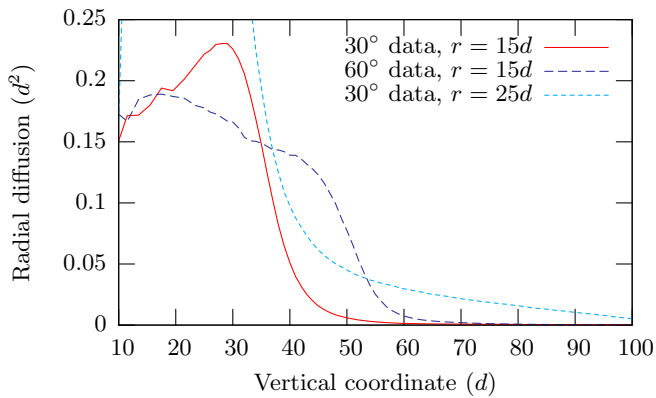}
	\caption{(Color online) Radial diffusion of particles about streamlines
	of the mean flow as a function of height, $z$, in both reactor
	geometries for pebbles starting at $z=110 d$ in an annulus of radius
	$r=15 d$, at the edge of the dynamic central column in MPBR. For the
	$30^\circ$ geometry, we also show data for pebbles near the wall at
	$r=25d$.}
	\label{diffz}
\end{figure}

We collected extensive statistics on how much pebbles deviate from the
mean-flow streamlines during drainage. Consistent with theoretical
concepts~\cite{spot-ses}, experiments have demonstrated that the dynamics are
strongly governed by the packing geometry, so that diffusion can most
accurately be described by looking at the mean-squared horizontal displacement
away from the streamline, as a function of the distance dropped by the pebble
(not time, as in molecular diffusion), regardless of the flow rate. Motivated
by the importance of quantifying mixing at the fuel/moderator interface in the
dynamic central column of MPBR, we focus on tracking pebbles passing through
$z=110d$ with $\abs{r-15d}<0.16d$. The variance of the $r$ coordinate of the
particles as they fall to different heights in $z$ can be calculated. From
this, we can determine the amount of radial diffusion, defined as the increase
in the variance of $r$ of the tracked particles from the variance at the
initial height.

The diffusion data for both reactor geometries is shown in figure \ref{diffz}.
We see that for large values of $z$ in the cylindrical part of the container,
the pebbles undergo essentially no diffusion; this is to be expected, since we
have seen that in this area the packing is essentially plug-like, and particles
are locked in position with their neighbors. However for lower values of $z$
the amount of radial spreading begins to increase, as the particles experience
some rearrangement in the region corresponding to converging flow. Note however
that the scale of this mixing is very small, and is much less than a pebble
diameter. For very small values of $z$, there is a decrease in the variance of
the radial coordinate, since the pebbles must converge on the orifice as they
exit the container. 

We applied a similar analysis for different initial values of $r$, and found
very similar results over the range $0<r<25d$. However, for particles close to
the container boundary, very different behavior is observed, as shown by the
third line in figure \ref{diffz} for particles with $\abs{r-25d}<0.10d$. In
this region, the particles undergo rearrangement, and this causes a (piecewise)
linear increase in the mean-squared displacement with distance dropped, which
corresponds to a constant local diffusion length. There is also evidence of a
sharp transition in the boundary-layer diffusion length, which increases
significantly as pebbles pass the corner into the converging-flow region of the
funnel.

\section{Packing Statistics}\label{sec:ordering}

\subsection{Pebble Volume Fraction} 

Pebble-bed experiments~\cite{goodling83,sederman01} and
simulations~\cite{dutoit02,ougouag05} of static sphere packings in cylinders
have revealed that there are local variations in porosity near walls, at the
scale of several pebble diameters, but there has been no such study of flowing
packings, averaging over dynamic configurations. Similar findings would have
important implications for helium flow in the core, since the local gas
permeability is related to the porosity~\cite{cohen81,vortmeyer83,white87}.

First, we study the distribution of local volume fraction (\% of volume
occupied by pebbles) throughout the container, averaged in time. (The porosity
is one minus the volume fraction.) Random close packing of spheres corresponds
to a volume fraction in the range 55\% - 63\%, while flows occur in a somewhat
more restricted range. The lower bound is approximately set by random loose
packing, where rigidity percolation sets \cite{onoda90}, while the upper bound
is near the jamming point \cite{ohern02} or the maximally random jammed state
\cite{torquato00}, where flow cannot occur.

\begin{figure}
	\centering
	\includegraphics[width=3.8cm]{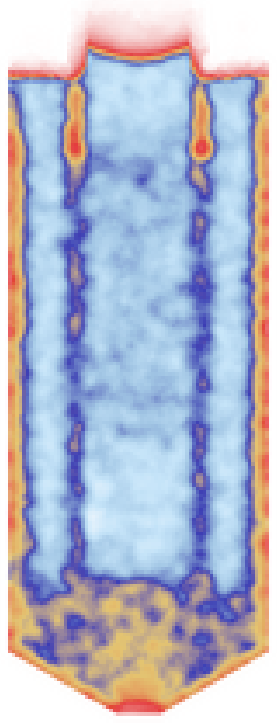} 
	\qquad
	\includegraphics[width=3.8cm]{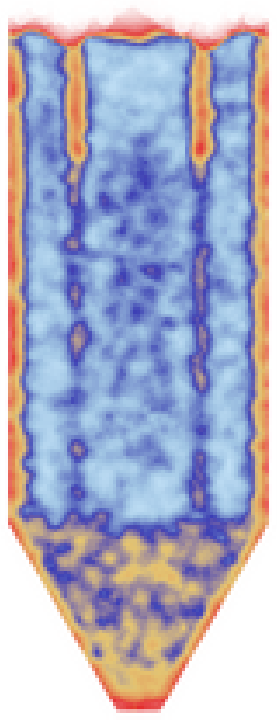}
	\caption{(Color online) Plots of local volume fraction
	($1-\textrm{porosity}$) in a vertical cross section for the thirty
	degree reactor geometry (left) and the sixty degree reactor geometry
	(right), calculated using a Voronoi cell method. The color scheme used
	is red 50\%, yellow 57\%, dark blue 60\%, cyan 63\%. High in the bulk
	of the container, the packing fraction is approximately 63\%, apart
	from in a small region of lower density at $\rin=14.5d$, corresponding
	to packing defects introduced by the guide ring. In both geometries a
	sharp reduction in density is observed in a region above the orifice,
	where particles in the parabolic flow region are forced to undergo
	local rearrangements.}
	\label{voro}
\end{figure}

\begin{figure}
	\centering
	\includegraphics[width=3.8cm]{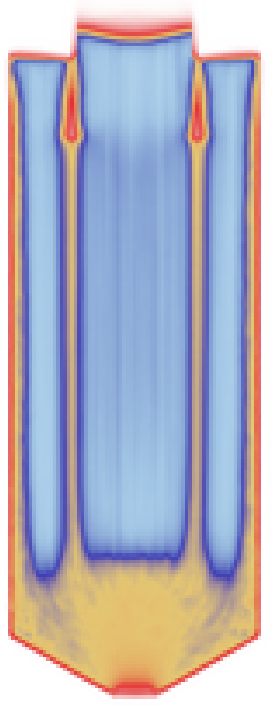} 
	\qquad
	\includegraphics[width=3.8cm]{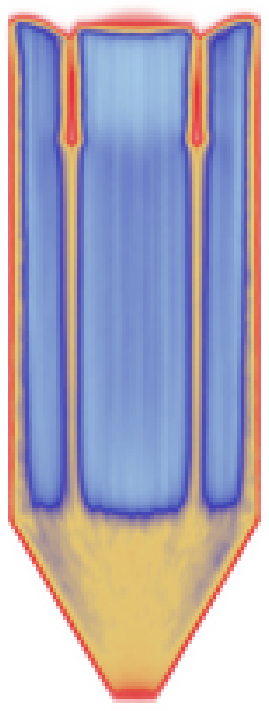}
	\caption{(Color online) Time-averaged plots of the local volume
	fraction, using the same color scheme as figure \ref{voro}.}
	\label{voro_av}
\end{figure}

The best way to determine the volume fraction on a local scale is to use a
Voronoi tessellation, which we compute with a novel efficient algorithm for
flows, to be described in detail elsewhere. The Voronoi tessellation uniquely
assigns a polygonal volume to each pebble, formed by intersecting the planes
bisecting the lines between different pebble centers. The local packing
fraction in a small region can then be found by taking the ratio of the
particle volume in that region to the ratio of the Voronoi volume. Such a
method can be used to define local density even down to the scale of a single
particle, but for this work we compute local density by averaging on a scale
of several particle diameters.

Figure \ref{voro} shows density snapshots for cross sections through the thirty
degree and sixty degree reactor geometries, based on computing the local
density at a particular point by averaging over the Voronoi densities of
particles within a radius of $2.2d$. Figure \ref{voro_av} shows density plots
over the entire flow of the data, but using a smaller averaging radius of
$0.8d$. Many interesting features are visible, which corroborate our other
results. High in the center of the container, we see that the local packing
fraction is mostly close to 63\%, suggesting that the plug-like region is in a
nearly jammed and rigid state. This is consistent with our earlier data showing
nearly uniform plug flow with no significant diffusion or mixing. 

We also observe two annular lines of lower density propagating down from the
guide ring, which form due to wall effects on the guide ring itself (see below)
and are advected downward. The fact that these subtle artifacts of the
guide-ring constraints are felt far down in the flow further demonstrates that
very little diffusion or shearing occurs in the upper region. There are also
similar lower-density regions along the walls, related to partial
crystallization described in more detail below.

It is also clear in both geometries, especially the $30^\circ$ model, that
there is a fairly sharp transition between the upper region of nearly rigid
plug flow and a less dense lower region of shear flow in the funnel. Similar
features are in the velocity profiles described above, but the transition is
much more sharp, at the scale of at most a few particles, for the local packing
fraction. These sudden variations in material properties and velocities are
reminiscent of shock-like discontinuities in Mohr-Coulomb plasticity theories
of granular materials~\cite{nedderman,schaeffer87}. It seems no such existing
theory can be applied to the reactor flows, but our results suggest that
plasticity concepts may be useful in developing a continuum theory of dense
granular flow~\cite{kamrin06}.

\subsection{Local Ordering and Porosity}

\begin{figure}
	\centering
	\input{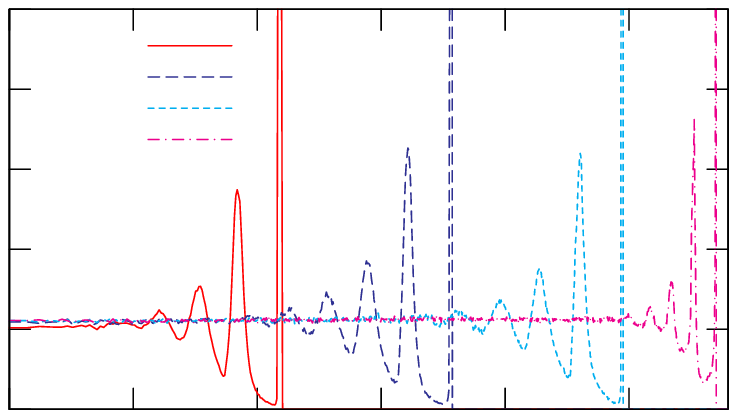}
	\input{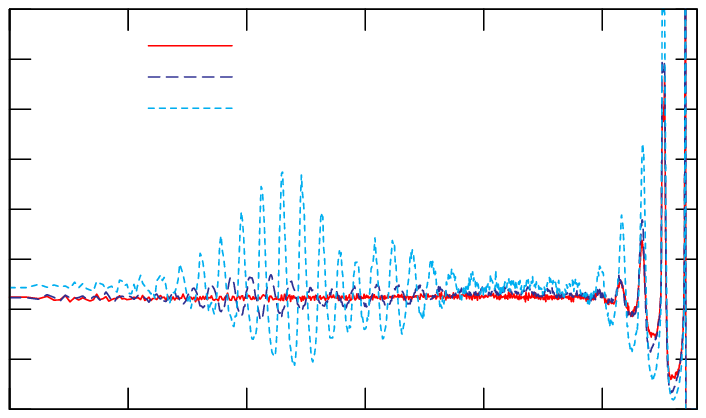}
	\caption{(Color online) Number density plots in the thirty degree
	reactor geometry for several low cross sections (a), and several high
	cross sections (b).}
	\label{n30}
\end{figure}

\begin{figure}
	\input{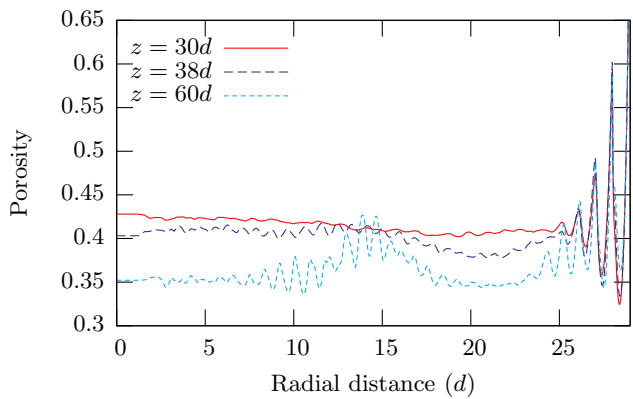}
	\caption{(Color online) Horizontal profiles of porosity at different
	heights in the $30^\circ$ reactor geometry.}
	\label{poro}
\end{figure}

As noted above, previous simulation studies of local ordering near walls have
focused on static packings in simplified cylindrical geometries (without the
funnel, outlet pipe, or guide ring)~\cite{dutoit02,ougouag05}, while we
compute average statistics for slowly flowing packings in realistic full-scale
reactor models. To take a closer look at ordering near walls, we study the
number density profile in horizontal slices at different heights. The container
is divided into bins in the same way as discussed previously and the number
density in a bin is obtained by counting the number of times a particle center
lies within that bin.

Figure \ref{n30}(a) shows a sequence of number density profiles for several low
values of $z$ in the thirty degree reactor geometry. At all four heights,
lattice effects are clearly visible and quite similar to those observed in
experiments~\cite{goodling83,sederman01} and other
simulations~\cite{dutoit02,ougouag05}. For the lowest three heights, these
peaks are roughly $\sqrt{3}d$ apart, corresponding to particles crystallized
against the conical wall, while for the highest value of $z$, these effects are
roughly $1d$ apart, due to particles being crystallized against the cylindrical
wall. The above graph also shows that in the middle of the container, no
lattice effects are present.

However, this situation changes dramatically higher up in the container, as
shown in Fig.~\ref{n30}(b). As $z$ increases from $30d$ to $60d$, the interior
of the packing goes from being disordered to having a strong radial ordering,
centered at around $z=12d$. The reason for this ordering is due to the presence
of the guide ring high in the container, which keeps the fuel and moderator
pebbles separate. The ring, placed at $\rin=14.5d$ in the container, creates
radial crystallization, which can then propagate very far downward, since the
packing is plug-like for most of the cylindrical part of the reactor. At much
lower heights, around $z=40d$, this radial ordering is broken, as the particles
are forced to reorganize once they enter the parabolic region of flow.

To make a direct connection with the modeling of gas flow, we show horizontal
slices of the porosity at different heights in figure \ref{poro}. The porosity
is measured here by intersecting the spheres with annular cylindrical bins to
compute the fraction of each bin volume not occupied by pebbles. The features
noted above appear in the porosity and alter the local permeability, which
enters continuum descriptions of helium gas flow in the
core~\cite{cohen81,vortmeyer83,white87}.

\section{Residence-Time Distribution}\label{sec:wait}

\subsection{Predictions of the Kinematic Model}

The statistical distribution of fuel burnup is closely related to the
distribution of pebble residence times in the reactor core, differing only due
to nonuniform sampling of the neutron flux profile. Since the upper pebble flow
is essentially a uniform plug flow, the distribution of residence times is the
same (up to a constant time shift) as the distribution of waiting times for
pebbles starting at a given height in the core to exit through the orifice, and
we concentrate on these distributions in this section. However, we conclude by
examining the residence times for particles to pass through the entire
container, to investigate the effects of the guide ring and the outer walls.

We have seen that there is very little pebble diffusion, so fluctuations in the
residence time are primarily due to hydrodynamic dispersion in the mean flow.
We have also seen that the Kinematic Model gives a reasonable description of
the mean flow profile in the conical funnel region, where most of the shear and
hydrodynamic dispersion occur. Therefore, we can approximate the residence-time
distribution by the distribution of times to travel along different streamlines
of the mean flow, starting from different radial positions, $r_0$, at a given
height $z_0$. Below we will compare such predictions, based on our numerical
solutions to the Kinematic Model, to our DEM simulations for the two reactor
geometries.

\subsection{An Analytical Formula}

We can obtain a simple, exact formula for the residence-time distribution in a
somewhat different geometry using the Kinematic Model, as follows. The
similarity solution to Eq.~(\ref{eq:veq}) for a wide, flat bottomed silo
draining to a point orifice at $z=0$ is
\begin{eqnarray}
u(r,z) &=& - \frac{Qr}{2bz^2} e^{-r^2/4bz}  \label{eq:up} \\
v(r,z) &=& \frac{Q}{bz} e^{-r^2/4bz} \label{eq:vp} 
\end{eqnarray}
where $u$ and $v$ are the radial (horizontal) and downward velocity components
and $Q$ is a constant proportional to the total flow rate through the orifice.
(This is just the classical Green function for the diffusion equation in two
dimensions, where $z$ acts like ``time''.) A slightly more complicated solution
is also possible for a parabolic silo, but let us focus on the simplest case of
Eqs.~(\ref{eq:up})-(\ref{eq:vp}), which is a good approximation for a wide
parabolic funnel, where the velocity near the walls is small, i.e. $R >
\sqrt{4bz_0}$. A more detailed analysis is not appropriate here, since a simple
analytical solution does not exist for the actual reactor geometry of a conical
funnel attached to straight cylinder.

For the flow field in Eqs.~(\ref{eq:up})-(\ref{eq:vp}), the trajectory of a
Lagrangian tracer particle along a streamline is given by
\begin{eqnarray}
\frac{dr}{dt} &=& u(r,z), \ \ \ r(t=0) = r_0 \\ 
\frac{dz}{dt} &=& -v(r,z), \ \ \ z(t=0) = z_0
\end{eqnarray}
Combining these equations and integrating, we find that the streamlines are
parabolae, $z/z_0 = (r/r_0)^2$, and that the residence time for a pebble
starting at $(r_0,z_0)$ is
\begin{equation}
\tau_0(r_0,z_0) = \frac{bz_0^2}{2Q} e^{r_0^2/4bz_0}.
\end{equation}

Now we consider pebbles that are uniformly distributed at a height $z_0$ in a
circular cross section of radius $R$ in the flow field
Eqs.~(\ref{eq:up})-(\ref{eq:vp}. The probability distribution for the residence
time of those pebbles is
\begin{eqnarray}
p(\tau|z_0,R) &=&
\int_0^R \delta(\tau - \tau_0(r_0,z_0)) \frac{2\pi r_0 dr_0}{\pi R^2}\\
&=& \left\{ \begin{array}{ll}
0 & \mbox{for } \tau < \tau_{min}(z_0) \\
4bz_0/R^2 \tau &
\mbox{for } \tau_{min} < \tau < \tau_{max} \label{eq:pkm} \\
0 & \mbox{for } \tau > \tau_{max}(z_0,R) 
\end{array} \right.
\label{eq:p}
\end{eqnarray}
where 
\begin{eqnarray}
\tau_{min}  &=& \tau_0(0,z_0) = \frac{bz_0^2}{2Q} \\
\tau_{max} &=& \tau_0(R,z_0) = \frac{bz_0^2}{2Q} e^{R^2/4bz_0}
\end{eqnarray}
Once again, this solution is strictly valid for an infinitely wide and tall
silo draining to a point orifice, and it is roughly valid for a parabolic
funnel, $ z/z_0 = (r/R)^2$, as an approximation of a conical funnel in the
actual reactor geometry. We can further approximate the effect of a nearly
uniform flow of speed $v_0$ to describe the upper cylindrical region by simply
adding $(z-z_0)/v_0$ to the residence time for a starting point $z>z_0$.

Although this analysis is for a modified geometry, we will see that it captures
the basic shape of the residence-time distributions from the DEM simulations in
a simple formula (\ref{eq:pkm}). The probability density is sharply peaked near
the shortest residence time, $\tau_{min}$, corresponding to pebbles near the
central axis traveling the shortest distance at the largest velocity. The
longer distance and (more importantly) the smaller velocity at larger radial
positions cause strong hydrodynamic dispersion, resulting a fat-tailed
residence-time density which decays like $1/t$, up to a cutoff $\tau_{max}$.

\subsection{Simulation Results}

For the DEM reactor simulations, we calculate the distribution of times it
takes for particles to drop from several different values of $z_0$, adding in a
weighting factor to take into account that shorter residence times are
preferentially observed in the data set.

\begin{figure}
	\centering
	\input{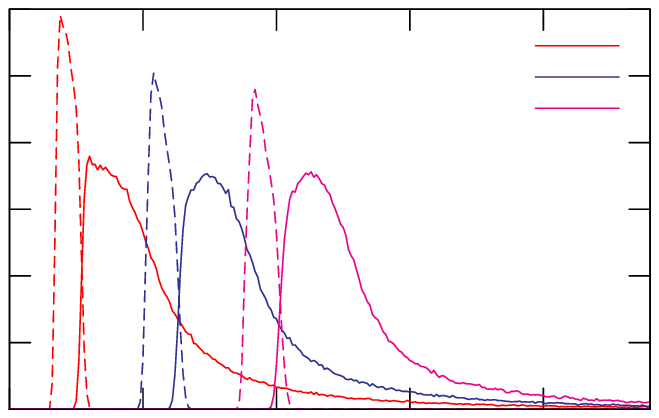}
	\input{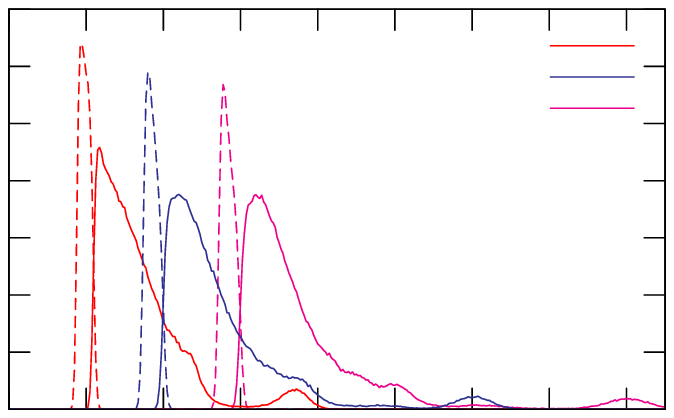}
	\caption{(Color online) Residence-time probability densities for the
	time it takes particles to drop from a specific height $z$ out of the
	container, for the thirty degree reactor geometry (a) and sixty degree
	reactor geometry (b) for fuel pebbles (solid lines) and for moderator
	pebbles (dashed lines).}
	\label{wait1}
\end{figure}

Since we are primarily interested in the radioactive burnup, we concentrate on
the residence times for the fuel pebbles, but for comparison, we also report
results for the moderator pebbles. Figure \ref{wait1}(a) shows the
residence-time probability densities for pebbles starting at $z=40d,55d,70d$ to
exit the container for the $30^\circ$ reactor geometry. The distributions for
the moderator pebbles are quite narrow, showing all particles exit over a short
time window. In contrast, the distributions for the fuel pebbles exhibit fat
tails, as expected qualitatively from the Kinematic Model approximation
(\ref{eq:pkm}) for a parabolic geometry. A closer analysis of the data confirms
that the longest waiting times are associated with pebbles passing close to the
walls, especially near the corner between the conical and cylindrical wall
sections, although there are no completely stagnant regions.

\begin{figure}
	\centering
	\input{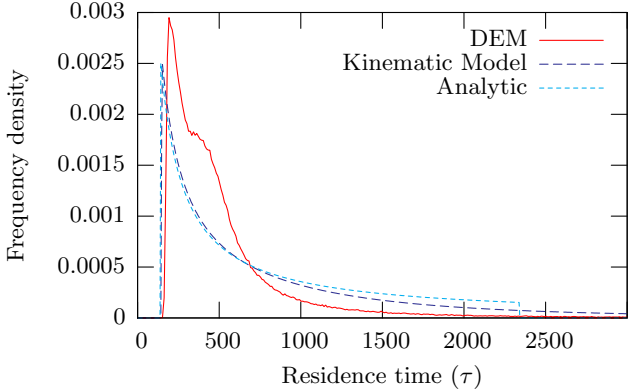}
	\caption{(Color online) Comparison of the residence time distributions
	between DEM simulation, numerical solution of the Kinematic Model, and
	the analytic formula.}
	\label{wait3}
\end{figure}

Figure \ref{wait1}(b) shows corresponding plots for the $60^\circ$ reactor
geometry. In general, the residence-time densities have similar shapes as for
the $30^\circ$ geometry, but they are much narrower and exhibit a small
secondary peak far into the tail. Examining movies shows that this extra peak
is due to a boundary layer of particles, roughly one-pebble thick, touching the
$60^\circ$ conical wall sliding down at a speed lower than the nearby bulk.
This extra source of hydrodynamic dispersion could not be easily captured by a
continuum model for the mean flow. A simple way to eliminate it would be to
replace add an outer annulus of moderator pebbles (controlled by another guide
ring at the top), which would flow more slowly along the walls, leaving the
fuel pebbles in a more uniform flow with smaller fluctuations. Another
possibility would be to reduce the wall friction, which makes the flow more
uniform, as discussed in the following section.

Figure \ref{wait3} investigates the accuracy of the Kinematic Model in
predicting the DEM residence-time distribution. The total residence-time
distribution for both fuel and moderator pebbles to exit the reactor from
$z=40d$ in the $30^\circ$ geometry is shown, and is compared with two
predictions from the Kinematic Model, one making use of the analytic
formula (\ref{eq:pkm}), and one making use of the numerical solution of the
velocity profile. We use of the value $b=2.5d$ and calibrate the total flow to
match the total flow from the DEM data. Both the numerical solution and the
analytic formula can roughly capture the overall shape of the DEM distribution,
although neither achieves a good quantitative agreement, particularly in the
tails. Since the analytic formula assumes all streamlines are parabolic, it
fails to take into account the slow-moving particles that stay close to the
wall, and it therefore predicts a cut-off in the residence time distribution
which is much shorter than some of the observed residence times in the DEM
simulation. The numerical solution of the Kinematic Model accounts for this
and provides a better match, although it is clear that a model correctly
accounting for the flow of pebbles near the container walls may be required
in order to achieve high accuracy.

\subsection{Residence times for the entire container}
We also considered the distribution of times for the particles to pass through
the entire container. While the flow in the upper part of the reactor is
essentially plug-like, boundary effects near the container walls and on the
guide ring can have an appreciable effect on the pebble residence times, which
we study here. Since it takes a long time for particles to pass through the
entire container we made use of the two extended data sets, consisting of 1,427
snapshots for the thirty degree geometry and 1,249 snapshots for the sixty
degree geometry.

\begin{figure}
	\input{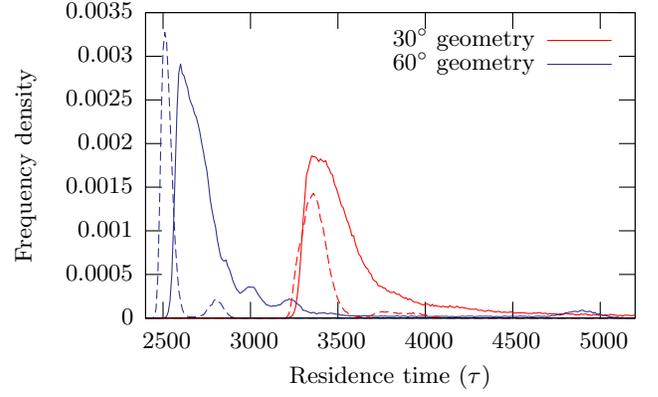}
	\caption{(Color online) Distribution of times to pass through the
	entire container for fuel pebbles (solid lines) and moderator pebbles
	(dashed lines).}
	\label{waitcomp}
\end{figure}

Figure \ref{waitcomp} shows the time distributions for pebbles to pass through
the entire container. Apart from a large positive time shift, the curves are
similar in form to those in Fig.~\ref{wait1}. However, for both geometries, we
see second small peaks in the distributions for the moderator pebbles,
corresponding to a slow-moving boundary layer of pebbles touching the guide
ring. The sixty degree curve for the fuel pebbles also exhibits several
undulations corresponding to multiple layers of pebbles crystallized against
the outer wall, each moving at different speeds.

\section{Wall friction}\label{sec:fric}

\begin{figure}
	\input{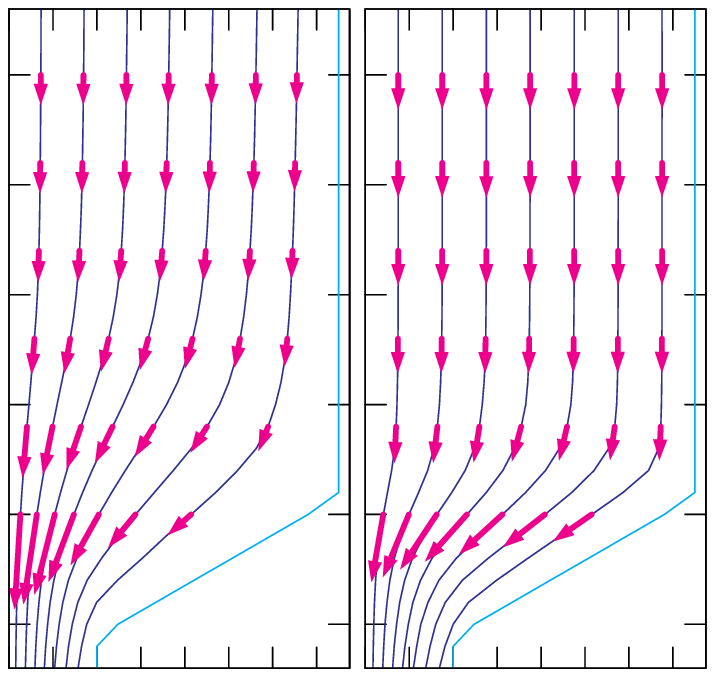}
	\caption{(Color online) Streamlines for the half-size, monodisperse
	geometries with wall friction (left) and without wall friction (right).
	Arrows are proportional to the velocity vectors in selected horizontal
	slices.}
	\label{streamfrict}
\end{figure}

\begin{figure}
	\centering
	\input{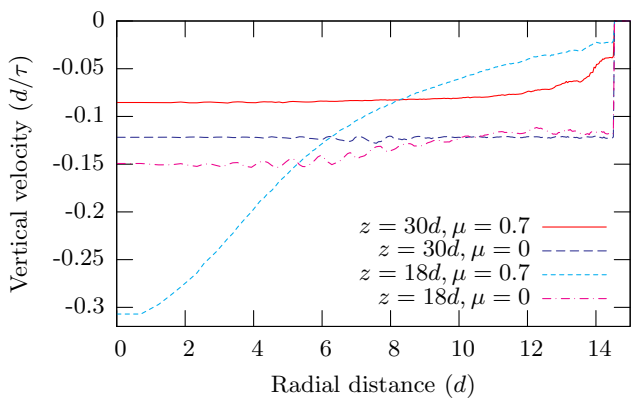}
	\caption{(Color online) Comparison of velocity profiles for simulations
	with and without wall friction for two different heights.}
	\label{fricv}
\end{figure}

\begin{figure}
	\centering
	\input{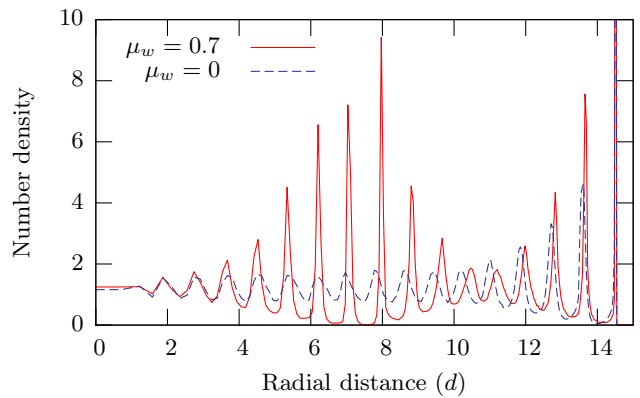}
	\caption{(Color online) Comparison of number density profiles at
	$z=60d$ for simulations with and without wall friction.}
	\label{nwall}
\end{figure}

The behavior of pebbles near the walls is of significant interest to reactor
design, and to look into this further, we investigated the effect of wall
friction by comparing two simulations runs in the half-size geometry, with
wall friction coefficients $\mu_w=0$ and $\mu_w=0.7$. All other aspects of
the simulation, including the interparticle interactions, were kept the same.

Figure \ref{fricv} shows a comparison of flow profiles for the two simulations
at two different heights. We see that the $\mu_w=0$ simulation results in a
significantly larger flow speed, with a mass flow rate of $104m\tau^{-1}$, as
opposed to $59.6m\tau^{-1}$ for $\mu_w=0.7$. As would be expected, removing
wall friction also removes the boundary layer of slower velocities at the wall,
creating an almost perfectly uniform velocity profile high in the reactor. This
also has the effect of increasing radial ordering effects, and we can see from
figure \ref{nwall} that the number density profile is more peaked close to the
wall. Figure \ref{nwall} also shows that the radial ordering created by the
guide ring is also significantly enhanced. While this is due in part to the
more plug-like flow allowing packing effects to propagate further down, it is
also due to the frictionless guide ring initially creating radial ordering.
Thus it may be possible to tune the material properties of the guide ring (or
the roughness of its walls) to enhance or reduce the radial ordering effects.

Removing wall friction also has the effect of increasing radial ordering
effects near the wall. Perhaps most surprisingly, removing wall friction
results in a significant alteration of the flow in the {\it interior} of the
packing, as shown by the two velocity profiles in figure \ref{fricv} for
$z=18d$. While both velocity profiles must converge upon the orifice, we see
that the velocity profile for the $\mu_w=0.7$ case is significantly more curved
than that for $\mu_w=0$. This also has the effect of preferentially speeding up
the relative flux of fuel pebbles: with wall friction, the fuel pebbles make up
$71.5\%$ of the total mass flux, but without wall friction, this increases to
$74.7\%$.

\section{Bidispersity}\label{sec:bi}

\subsection{The Bidisperse PBR Concept}

\begin{figure}
	\includegraphics[width=3in]{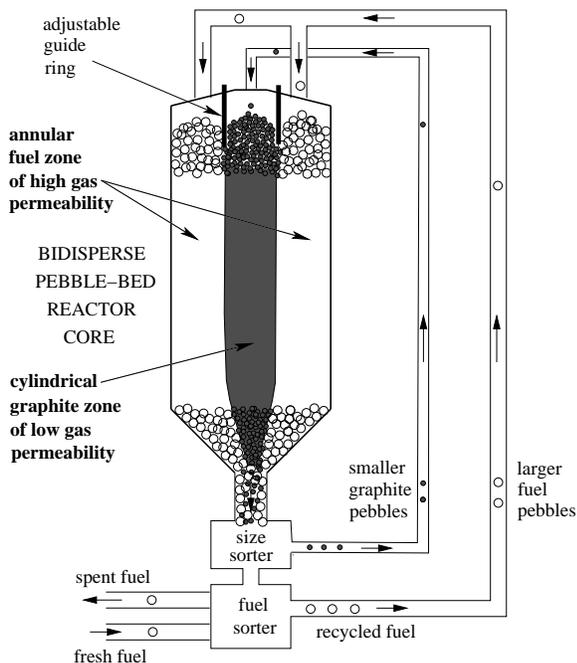}
	\caption{Schematic diagram of the pebble flow in a bidisperse MPBR
	design.}
	\label{bidisperse}
\end{figure}

The two-pebble design of MPBR with a dynamic central moderator column has
various advantages over a solid graphite central column (as in the revised PBMR
design). For example, it flattens the neutron flux profile, while preserving a
very simple core vessel without any internal structures, which would be
subjected to extreme radiation and would complicate the granular flow. It also
allows the widths of the moderator column and fuel annulus to be set ``on the
fly'' during reactor operation, simply by adjusting the guide ring at the top.

A drawback of the dynamic moderator column, however, is its porosity, which
allows the passage of the helium-gas coolant, at the highest velocity (along
the central axis). To improve the thermal efficiency and power output, it would
be preferable to focus the gas flow on the fuel annulus and the interface with
the moderator column, where the most heat is generated. This is automatically
achieved with a solid graphite central column, but there is a very simple way
to shape the gas flow in a similar way with a dynamic column, while preserving
its unique advantages.

The idea is to make the graphite moderator pebbles in the central column
smaller than the fuel pebbles in the outer annulus, as shown in
Fig.~\ref{bidisperse}. (This also helps with sorting of fuel and moderator
pebbles as they exit the core.)  In standard continuum models of flow in porous
media~\cite{cohen81,vortmeyer83,white87}, the permeability of the packing
scales with the square of the pebble diameter (or pore size), so reducing the
diameter of the moderator pebbles can greatly reduce the gas flow (e.g. by a
factor of four for half-diameter pebbles). This argument holds everywhere that
the packing is statistically the same, in the monodisperse packings of the fuel
annulus and the moderator column, which have the same porosity. At the
interface between the two regions, we have seen in Figures \ref{voro} and
\ref{poro} that the porosity is enhanced for a monodisperse core due to the
guide ring, although a bidisperse interface will have different structure.

\begin{figure}
\includegraphics[width=3.1in]{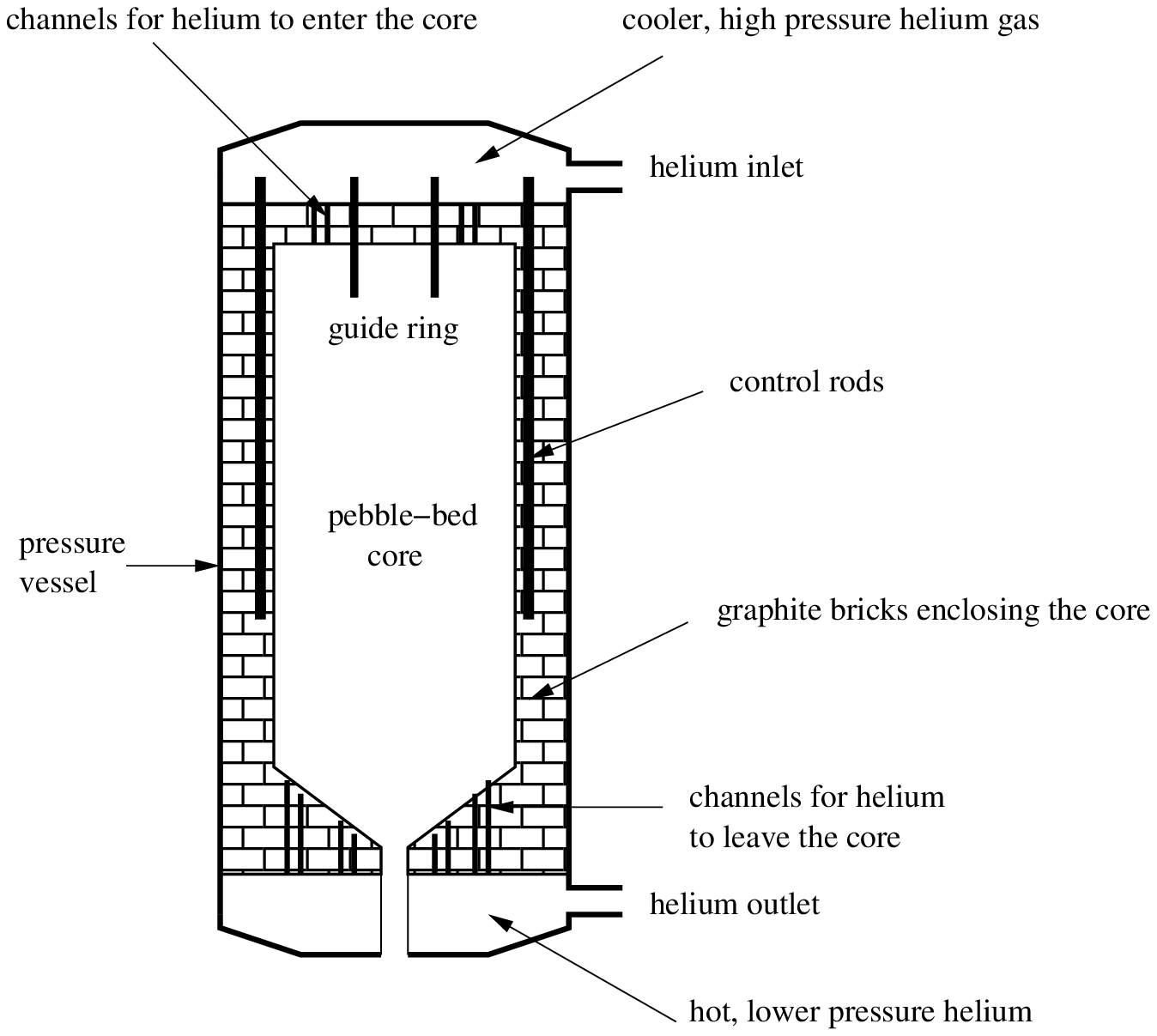} \\
\ \\
\includegraphics[width=0.9in]{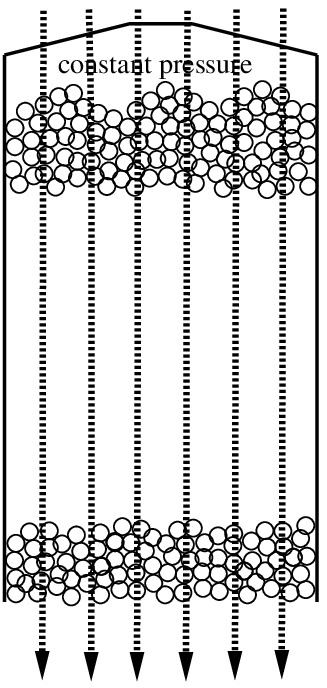}\  \nolinebreak
\includegraphics[width=0.9in]{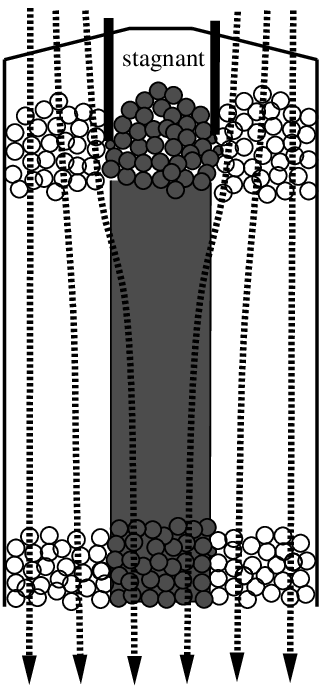}\  \nolinebreak
\includegraphics[width=0.9in]{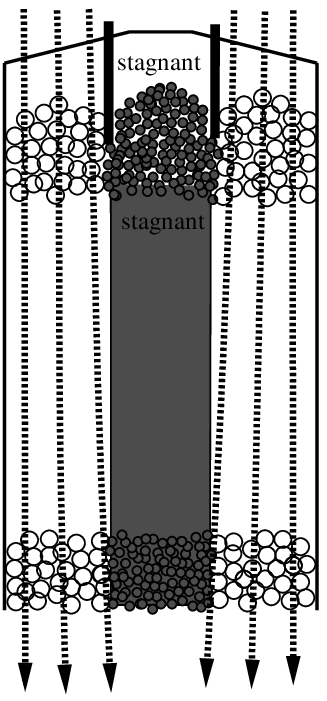}
\caption{Sketches of the helium-gas coolant system (top) and the
  flows a basic PBR core (left), a monodisperse MPBR where gas is
  introduced only in the fuel annulus outside the guide ring (center),
  and a bidisperse MPBR where the dynamic moderator column has much
  lower permeability due to smaller pebble size (right).}
\label{helium}
\end{figure}

In any case, it is clear that the bidisperse core will focus the coolant flow
away from the moderator column and onto the fuel annulus, as shown in Figure
\ref{helium}. In most PBR designs, high-pressure helium gas is introduced from
a reservoir above the core, through holes in the graphite bricks which make up
the core vessel. The gas then flows through the core and exits through holes in
the graphite bricks in the conical funnel to another reservoir at ``very high''
temperature ($\approx 950^\circ$C). In MPBR, the gas can be introduced only
outside the guide ring, which focuses the gas flow on the fuel annulus down to
a distance comparable to the radius of the guide ring. With a significant
reduction in permeability of the central column in the bidisperse core, the gas
flow can be focused almost entirely on the fuel annulus and the interfacial
region (where the heat generation is maximal).

\subsection{Simulation Results}

The only question regarding the feasibility of the bidisperse core is the
stability of the central column over time and the possibility of enhanced
diffusion of the small moderator pebbles into the annulus of larger fuel
pebbles. In other systems, such as rotating drums \cite{newey04,khan04,khan05},
vibrated buckets \cite{shinbrot98,shinbrot04}, and draining
silos~\cite{samadani99}, bidisperse granular materials display a tendency to
segregate (rather than mix) during dynamics, but there is currently no general
theory which could be applied to our reactor geometry. Therefore, our DEM
simulations provide a useful means to address this important question.

\begin{figure}
	\centering
	\includegraphics[width=2.5cm]{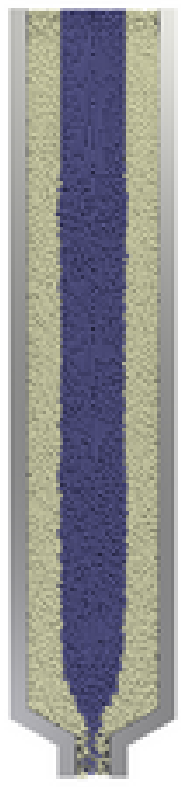}\quad
	\includegraphics[width=2.5cm]{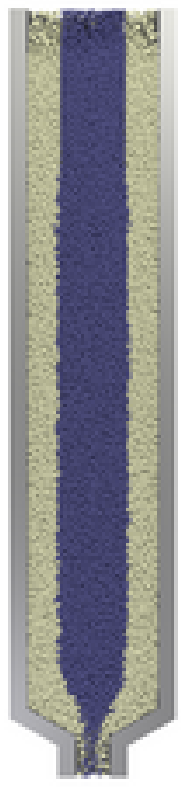}\quad
	\includegraphics[width=2.5cm]{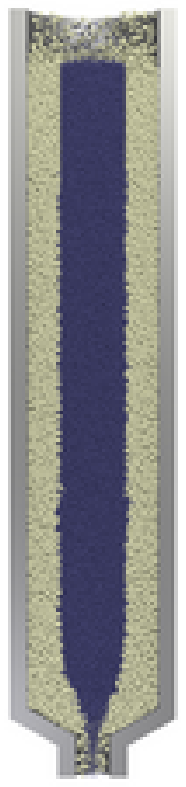}
	\caption{(Color online) Snapshots of vertical cross-sections for the
	bidisperse simulations. From left to right, the moderator pebbles have
	diameters $1d$, $0.8d$, and $0.5d$ while the fuel pebbles are of
	constant size $1d$.}
	\label{bidiag}
\end{figure}

\begin{figure}
	\centering
	\input{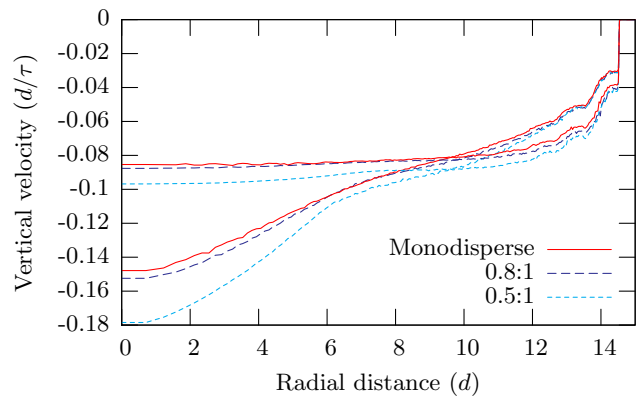}
	\caption{(Color online) Comparison of velocity profiles for the three
	bidisperse simulations. The three flatter curves are calculated at
	$z=30d$ in the plug-like flow region while the other three were taken
	at $z=22d$ in the parabolic flow region.}
	\label{bivel}
\end{figure}

Figure \ref{bidiag} shows snapshots of vertical cross sections for the three
different bidisperse simulations that were run in the half-size geometry. As
shown in the diagram, the central column remains stable and coherent in all
three cases, and very little mixing between the two types of pebbles is
visible. Figure \ref{bivel} shows a comparison of the velocity profiles from
the three simulations for two different heights. It is reassuring to see that
the bidisperse simulations do not significantly differ from the monodisperse
simulation, although we do see a slightly higher overall flow rate in the
bidisperse systems: we see total mass flow rates of $59.6m\tau^{-1}$,
$60.8m\tau^{-1}$, and $65.0m\tau^{-1}$ for the monodisperse, 0.8:1, and 0.5:1
simulations respectively.

The velocity profiles are slightly more curved in the bidisperse central core;
this is particularly apparent in the 0.5:1 simulation. This leads to a small
cusp in the velocity profile near the interface between the two types of
particles which may lead to adverse mixing effects. The faster flow also leads
to a significantly larger turnaround of the moderator pebbles. In the
monodisperse system, the moderator pebbles comprise $28.5\%$ of the total mass
flux, but this is increased to $31.7\%$ in the 0.8:1 bidisperse simulation, and
$42.6\%$ in the 0.5:1 bidisperse simulation.

\begin{figure}
	\centering
	\input{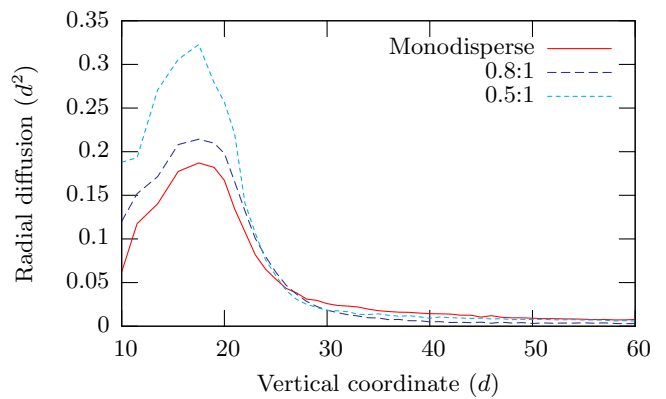}
	\caption{(Color online) Comparison of particle diffusion for the three
	bidisperse simulations.}
	\label{bidiff}
\end{figure}

To investigate the amount of mixing of the central column, we used a technique
similar to that described in section \ref{sec:diff}. At $z=110d$ all moderator
particles with $r>8d$ are marked, and their radial diffusion is then calculated
as a function of $z$. The results are shown in figure \ref{bidiff}: in the
cylindrical section of the packing, there is very little difference between the
three simulations, but in the area of convergent flow, we see that bidispersity
leads to significantly more mixing. However, even for the 0.5:1 simulation, the
scale of diffusion is still smaller than a single particle diameter, and
essentially the central column remains stable. 

Due to computational limitations, we were unable to investigate smaller size
ratios in the reactor geometries, so we carried out simulations in a smaller
container with a 0.3:1 size ratio and found dramatically different behavior:
During drainage, the central column became unstable, and the small particles
penetrated many particle diameters into the packing of larger particles. We
expect that there is a fundamental crossover in behavior simply due to geometry
of amorphous packings, when the moderator pebbles become small enough to pass
through the gaps between the densely packed fuel pebbles. An in-depth study of
this phenomenon remains a subject of future work. For now, we can safely
recommend a diameter ratio of 0.5:1, which reduces the dynamic central column's
permeability by a factor of four without introducing any significant diffusion
of moderator pebbles into the fuel annulus.

\section{Conclusions}
\label{sec:conc}

\subsection{Pebble-Bed Reactor Core Design}

Using DEM simulations, we have analyzed many aspects of granular flow in
pebble-bed reactor cores of direct relevance for design and testing. We close
by summarizing some key conclusions.

The mean flow profile exhibits a smooth transition from a nearly uniform plug
flow in the upper cylindrical region to a nonuniform, converging flow in the
lower funnel region, consistent with recent
experiments~\cite{choiexpt,kadak04}. There are no stagnant regions in the
$30^\circ$ and $60^\circ$ conical funnels considered in this study, although
the flow is slower near the corner at the top of the funnel, especially in the
former case. Moreover, the wider $30^\circ$ funnel has a boundary mono-layer of
slower pebbles partially crystallized on the wall.

The only available continuum theory for such flows, the simple Kinematic
Model~\cite{lit58,mullins72,nedderman79,nedderman}, gives a reasonable
qualitative picture of the flow profiles, although it cannot capture discrete
boundary-layer effects. As in other experiments on similar
geometries~\cite{choiexpt}, the Kinematic Model does not quantitatively predict
the dependence of the flow profile on geometry. We suggest that it be used to
get a rough sense of the flow profile for a given core geometry prior to (much
more computationally expensive) DEM simulations and/or experiments.
 
We have quantified the degree of pebble mixing in the core. Although there is
some horizontal diffusion in the funnel region, pebbles depart from the
streamlines of the mean flow by less than one pebble diameter prior to exiting
the core. 

We have demonstrated that the ``mixing layer'' between the central moderator
column and the outer fuel annulus, which appears in prior
models~\cite{gougar02}, can be reduced to the thickness of one pebble diameter
by separating moderator and fuel pebbles with a guide ring at the ceiling (to
eliminate mixing by surface avalanches), consistent with experiments on MPBR
models~\cite{kadak04}. We conclude that the dynamic central column of moderator
pebbles is a sound concept, which should not concern regulators.

We have constructed Voronoi tessellations of our flowing packings to measure the
profile of volume fraction (or porosity) and found some unexpected features
which would affect coolant gas flow through the core. The bulk of the core, in
the plug-flow region of the upper cylinder, has a volume fraction near the
jamming point ($63$\%), but there is a sharp transition to less dense packings
($55-60$\%) in the funnel region, due to shear dilation. We also observe lower
volume fractions in this range at the moderator/fuel interface in the upper
cylinder, below the guide ring, and lower volume fractions ($50-55$\%) against
the walls. These narrow regions of increased porosity (and thus, increased
permeability) would allow faster helium gas flow.

We have also studied local ordering in the flowing packings and find evidence
for partial crystallization within several pebble diameters of the walls,
consistent with previous experiments~\cite{goodling83,sederman01} and
simulations~\cite{dutoit02,ougouag05}. Such ordering on the walls of the guide
ring, then advected down through the core, is responsible for the increased
porosity of the moderator/fuel interface.

We have varied the wall friction in our DEM simulations and observe that it can
affect the mean flow, even deep into the bulk. Reducing the wall friction
increases radial ordering near the walls and makes the flow profile more
uniform.

Since diffusion is minimal, the probability distribution of pebble residence
times is dominated by advection in the mean flow. Therefore, we have made
predictions using the Kinematic Model, numerically for the conical-funnel
reactor geometries, and analytically for a wide parabolic funnel. The model
predicts a fat-tailed ($\sim 1/t$) decay of the residence-time density due to
hydrodynamic dispersion in the funnel region.

Our DEM simulations predict that the $60^\circ$ conical funnel results in a
narrower residence-time distribution than the $30^\circ$ funnel, which has more
hydrodynamic dispersion. The steeper $60^\circ$ funnel also exhibits a boundary
layer of slower, partially crystallized pebbles near the wall which lead to an
anomalous bump far in the tail of residence-time distribution. These results
have important implications for non-uniformity in the burnup of fuel pebbles.

We have introduced the concept of a bi-disperse core with smaller moderator
pebbles in the dynamic central column than in the outer fuel annulus, in order
to focus the helium gas flow on the fuel. Our DEM simulations demonstrate that
there is negligible pebble mixing at the interface for diameter ratios as small
as 0.5:1, for which the permeability of the moderator column is reduced by a
factor of four. We conclude that the bidisperse MPBR design is sound and will
produce a stable moderator-pebble column of greatly reduced gas permeability.

A natural next step would be to combine our full-scale DEM model for the pebble
flow with existing computational approaches to reactor core
physics~\cite{terry02,gougar02}, which rely on pebble flow as an empirical
input. More accurate studies of gas flow in the core could also be done,
starting from our complete pebble packings, or the average quantities such as
the porosity. With such computational tools, one should be able to reliably
test and develop new reactor designs.

\subsection{Basic Physics of Dense Granular Flow}

We have noted a number of favorable comparisons between our simulations and
experiments in similar
geometries~\cite{kadak04,choiexpt,goodling83,sederman01}, which provides
further validation of the Discrete-Element Method as a realistic means of
simulating granular materials. As such, it is interesting to consider various
implications of our results for the theories of dense granular flow, since the
simulations probe the system at a level of detail not easily attained in
experiments.

Our conclusions about the Kinematic Model are similar to those of a recent
experimental study~\cite{choiexpt}: The model describes the basic shape of the
flow field in the converging region, but fails to predict the nearly uniform
plug flow in the upper region with vertical walls or the precise dependence on
the funnel geometry. It also cannot describe boundary-layers due to partial
crystallization near walls or incorporate wall friction, which we have shown to
influence the entire flow profile.

On the other hand, there is no other continuum model available for dense silo
drainage, except for Mohr-Coulomb plasticity solutions for special 2d
geometries, such as a straight 2d wedge without any corners~\cite{nedderman},
so it is worth trying to understand the relative success of the Kinematic Model
for our 3d reactor geometries and how it might be improved. A cooperative
microscopic mechanism for random-packing dynamics, based on ``spots'' of
diffusing free volume, has recently been proposed, which yields the mean flow
of the Kinematic Model as the special case of independent spot random walks
with uniform upward drift from the orifice (due to gravity)~\cite{spot-ses}.
Under the same assumptions, the Spot Model has also been shown to produce
rather realistic simulations of flowing packings in wide silos (compared to DEM
simulations)~\cite{ssim}, where the Kinematic Model is known to perform
well~\cite{tuzun79,medina98a,samadani99,choi04}. This suggests that some
modification of the spot dynamics, such as spot interactions and/or nonuniform
properties coupled to mechanical stresses, and an associated modification of
the Kinematic Model in the continuum limit, may be possible to better describe
general situations.

From a fundamental point of view, perhaps the most interesting result is the
profile of Voronoi volume fraction (or porosity) in our flowing random packings
in Figure~\ref{voro}. Although the mean velocity in Figure~\ref{stream} shows a
fairly smooth transition from the upper plug flow to the lower converging flow,
the volume fraction reveals a sharp transition (at the scale of $1-3$
particles) from nearly jammed ``solid'' material in the upper region (63\%) to
dilated, sheared ``liquid'' material in the lower region (57-60\%). The
transition line emanated from the corners between the upper cylinder and the
conical funnel. We are not aware of any theory to predict the shape (or
existence) of this line, although it is reminiscent of a ``shock'' in the
hyperbolic equations of 2d Mohr-Coulomb plasticity~\cite{nedderman}.

Our measurements of diffusion and mixing provide some insights into statistical
fluctuations far from equilibrium. Consistent with the experiments in wide
quasi-2d silos~\cite{choi04}, we find that diffusion is well described
geometrically as a function of the distance dropped, not time (as in the case
of thermal molecular diffusion). As a clear demonstration, there is essentially
no diffusion as pebbles pass through the upper core, until they cross the
transition to the funnel region, where the diffusion remains small (at the
scale of one pebble diameter) and cooperative in nature. The behavior in the
funnel is consistent with the basic Spot Model~\cite{spot-ses}, but a
substantial generalization would be needed to describe the transition to the
upper region of solid-like plug flow, perhaps using concepts from plasticity
theory~\cite{kamrin06}.

We view silo drainage as a fundamental unsolved problem, at least as
interesting and relevant for applications as Couette shear cells, which have
been received much more attention in physics. The challenge will be to find a
single theory which can describe both shear cells and silo drainage. Our
results for pebble-bed reactor geometries may provide some useful clues.

\section{Acknowledgements}
This work was supported by the U. S. Department of Energy (grant
DE-FG02-02ER25530) and the Norbert Weiner Research Fund and the NEC Fund at
MIT. Work at Sandia was supported by the Division of Materials Science and
Engineering, Basic Energy Sciences, Office of Science, U. S. Department of
Energy. Sandia is a multiprogram laboratory operated by Sandia Corporation, a
Lockheed Martin Company, for the U.S. Department of Energy's National Nuclear
Security Administration under contract DE-AC04-94AL85000.

\appendix
\section{Numerical Solution of the Kinematic Model}
\label{appendix_kin}

In the Kinematic Model for drainage the vertical downward velocity
$u$ in the container is assumed to follow a diffusion equation of the form
\[
\frac{\p v}{\p z} = b \nabla^2_\perp v
\]
where $\nabla^2_\perp$ is the horizontal Laplacian. By exploiting the axial
symmetry, $v$ can be treated as a function of $z$ and $r$ only. In cylindrical
coordinates the Laplacian is
\begin{eqnarray*}
\frac{\p v}{\p z} &=& b \frac{1}{r} \frac{\p}{\p r} \left(r \frac{\p v}{\p r}\right) \\
&=& b \frac{\p^2 v}{\p r^2} + b \frac{1}{r}\frac{\p v}{\p r}.
\end{eqnarray*}
The radial velocity component is given by
\[
u= b\frac{\p v}{\p r}
\]
and by enforcing that the velocity field at the wall must be tangential to the
wall, we can obtain boundary conditions for solving $v$.

To solve the above equation in a cylinder is straightforward, since we can make
use of a rectangular grid. The boundary condition reduces to $v_r=0$ at the
wall. However, to solve this equation in the reactor geometry, we must also
consider the complication of the radius of the wall, $R$, being a function of
$z$. To ensure accurate resolution in the numerical solution of $v$ at the
wall, we introduce a new coordinate $\lambda=r/R(z), \eta=z$, which then allows
us to solve for $u$ over the range $0<\lambda<1$. Under this change of
variables, the partial derivatives transform according to
\begin{eqnarray*}
\frac{\p}{\p r} &=& \frac{1}{R(\eta)} \frac{\p}{\p \lambda} \\
\frac{\p}{\p z} &=& \frac{\p}{\p y}
- \frac{\lambda R'(\eta)}{R(\eta)} \frac{\p}{\p \lambda}.
\end{eqnarray*}
In the transformed coordinates
\[
R^2 v_\eta = \frac{b}{\lambda} v_\lambda + b v_{\lambda\lambda} + \lambda R R' v_\lambda.
\]
To ensure differentiability at $r=0$, we use the boundary condition
\begin{equation}
\label{zerocond}
\left.
\frac{\p v}{\p \lambda}\right|_{\lambda=0} = 0,
\end{equation}
and by ensuring zero normal velocity at the wall we find that
\begin{equation}
\label{onecond}
\left.
\frac{\p v}{\p \lambda}\right|_{\lambda=1} = -\frac{v R' R}{b}.
\end{equation}
To numerically solve this partial differential equation, we make use of the
implicit Crank-Nicholson integration scheme. We write
$v^n_j=v(j\Delta\lambda,n\Delta \eta)$, and solve in the range $j=0,1,\ldots,N$
where $N={\Delta\lambda}^{-1}$. Away from the end points, the Crank-Nicholson
scheme tells us that
\begin{eqnarray*}
\frac{v_j^{n+1} - v_j^n}{\Delta \eta} &=&
\frac{b}{2 \Delta \lambda^2 R^2} \big(v_{j+1}^{n+1}-2v_j^{n+1}+v_{j-1}^{n+1}+
v_{j+1}^n\\
&&-2v_j^n+v_{j-1}^n\big) + \left(\frac{b}{4j\Delta\lambda^2 R^2} + \frac{jR'}{4R}\right)\\
&&\times\big(v_{j+1}^{n+1}-v_{j-1}^{n+1}+v_{j+1}^n-v_{j-1}^n\big),
\end{eqnarray*}
where all references to $R$ and $R'$ are evaluated at $\eta=\Delta \eta
(j+\frac{1}{2})$. If $j=0$, then by reference to equation \ref{zerocond}, we
find that
\[
\frac{v_0^{n+1}-v_0^n}{\Delta \eta} = 
\frac{b}{\Delta \lambda^2 R^2} \left(v_1^{n+1}-v_0^{n+1}+v_1^n-v_0^n\right).
\]
Similarly, for $j=N$, by reference to equation \ref{onecond}, we see that
effectively
\[
\frac{v_{N+1}^n-v_{N-1}^n}{2\Delta \lambda} = -\frac{v_N^n R' R}{b} 
\]
and hence
\begin{eqnarray*}
\frac{v_N^{n+1} - v_N^n}{\Delta \eta} &=&
\frac{b}{\Delta \lambda^2 R^2} \left(v_{N-1}^{n+1}-v_N^{n+1}+
v_{N-1}^n-v_N^n\right)\\
&&- \left( \frac{(2N+1)R'}{2 R} + \frac{R'^2}{2b}\right)\left(v_N^{n+1}+v_N^n\right).
\end{eqnarray*}
If we write $\vec{v}^n = (v_0^n, v_1^n, \ldots, v_N^n)^T$, then the above
numerical scheme can be written in the form $S \vec{v}^{n+1} = T \vec{v}^n$
where $S$ and $T$ are tridiagonal matrices; this system can be efficiently
solved by recursion in $O(N)$ time. The above scheme was implemented in C++,
and gives extremely satisfactory results, even with a relatively small number
of gridpoints. 

\bibliography{granular}

\end{document}